# Evidence of Photocatalytic Dissociation of Water on TiO$_2$ with Atomic Resolution

Shijing Tan, Yongfei Ji, Yang Wang, Jin Zhao, Aidi Zhao, Bing Wang*, Yi Luo, Jinlong Yang, and J. G. Hou*

Hefei National Laboratory for Physical Sciences at the Microscale, University of Science and Technology of China, Hefei, Anhui 230026, P.R. China. *Correspondence should be addressed to these authors: bwang@ustc.edu.cn and jghou@ustc.edu.cn

**Abstract**: Photocatalytic water splitting reaction on TiO$_2$ surface is one of the fundamental issues that bears significant implication in hydrogen energy technology and has been extensively studied. However, the existence of the very first reaction step, the direct photo-dissociation of water, has been disregarded. Here, we provide unambiguously experimental evidence to demonstrate that adsorbed water molecules on reduced rutile TiO$_2$(110)-1×1 surface can be dissociated under UV irradiation using low temperature scanning tunneling microscopy. It is identified that a water molecule at fivefold coordinated Ti (Ti$_{5c}$) site can be photocatalytically dissociated, resulting in a hydroxyl at Ti$_{5c}$ and another hydroxyl at bridge oxygen row. Our findings reveal a missing link in the photocatalytic water splitting reaction chain, which greatly contribute to the detailed understanding of underlying mechanism.

The discovery of photoelectrochemical water splitting on TiO$_2$ electrode *(1)* has triggered intensive study on its reaction mechanism *(2,3)*, aiming for the increase of its efficiency and the design of better photocatalysts. This seemingly simple reaction actually consists of complicated reaction steps that have not been fully understood. It has long been assumed that the hydroxyls on the TiO$_2$ surface play the key role in the whole reaction chain *(4,5)*. How the reactions initiated by the hydroxyls under the light irradiation has been the subject of many theoretical and experimental studies over the last decades *(6-8)*. However, less attention has been paid to the direct dissociation of water under light irradiation, the very initial step of the photocatalytic reaction. It has been observed that the water molecules adsorbed on bridge-bonded oxygen vacancies (BBO$_V$) can dissociate into pairs of hydroxyl groups (OH$_b$) residing on bridge-bonded oxygen (BBO) rows on reduced TiO$_2$ (110)-1×1 surface *(9-13)*. The study of the laser-induced fluorescence (LIF) spectroscopy indicated the existence of another type of hydroxyls, namely •OH$_t$ located at the fivefold coordinated Ti (Ti$_{5c}$) sites *(14)*, which should be a key intermediate for the successive reactions in photocatalytic water splitting *(2,7,8)*, but the origin of •OH$_t$ is not known. To identify the reaction process that generates such specie requires microscopic studies with atomic resolution, but to the best of our knowledge, there is no such attempt has ever been made. We have carried out a comprehensive experimental study on the direct photocatalytic dissociation of water on TiO$_2$ surface to address this important issue. We have chosen a single crystal rutile TiO$_2$ (110)-1×1 surface under UHV conditions with low temperature scanning tunneling microscopy (STM) to highlight the fundamental physics and chemistry involved in the processes. We have found that under UV irradiation, the •OH$_t$ can indeed be generated by the photocatalytic dissociation of the adsorbed water, which strongly suggest that direct photocatalytic dissociation of water is an important initial step in the whole water splitting reaction chain.

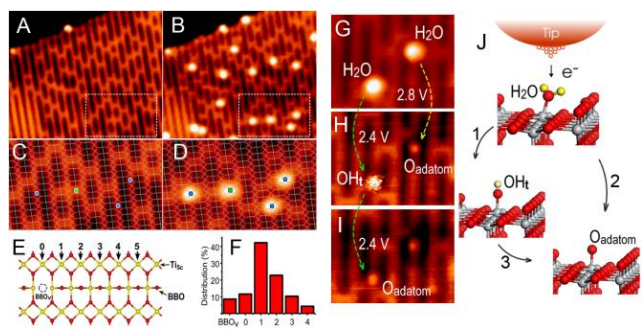

**FIG. 1.** Adsorption of water molecules and tip-induced dissociation on TiO$_2$(110)-1×1 surface. (A) and (B) Images of TiO$_2$(110)-1×1 before and after in situ H$_2$O adsorption at 80 K. (Size: 13.3 ×11.0 nm$^2$). (C) and (D) Magnified images superposed with structural model of TiO$_2$ surface, showing detailed H$_2$O adsorption sites (6.1×3.6 nm$^2$). Circles: adsorption sites of water at Ti$_{5c}$, rectangles: adsorption sites of water at BBO$_V$. (E) Defined specific Ti$_{5c}$ sites with respect to the BBO$_V$. (F) Distribution of water adsorption at the different sites. (G)-(I) STM images (Size: 3.9×3.0 nm$^2$) of the tip-induced water dissociation to produce OH$_t$ or oxygen adatom under different applied bias voltages. (J) Schematic drawings showing structural models of different processes of the tip-induced water dissociation. Imaging conditions: 1.0 V, 10 pA, 80 K.

We characterized the adsorption behavior of water molecules at 80 K, as shown in **Fig. 1** (See detailed experimental methods in Supporting Online Materials). On the clean $TiO_2$ surface with the $BBO_V$ concentration of about 0.1 ML (1 ML = $5.2 \times 10^{14}$ $cm^{-2}$) (**Fig. 1A**), we had water dosed and obtained the water coverage of about 0.02 ML (**Fig. 1B**, within the same area as in **Fig. 1A**). It is observed that the adsorbed water molecules are quite immobile at this temperature, showing a different behavior from those at elevated temperatures *(15)*. The specific adsorption sites of water molecules are identified by superposing the structural model of the surface (**Fig. 1C** and **1D**). By counting several thousands of adsorbed water molecules, the distribution of the adsorbed molecules is obtained and plotted in **Fig. 1F** with respect to the defined sites in **Fig. 1E**. Most of the adsorbed water molecules appear at the $Ti_{5c}$ sites, and only 8.5% water molecules at the $BBO_V$s. Except for the relative high adsorption percentage of water at the $BBO_V$s, the adsorption distribution of water is somewhat similar to the CO adsorption *(16)*, which may reflect the electronic nature of the surface *(17)*. It is observed that the adsorbed water molecules at the $BBO_V$s may undergo dissociation into pairs of $OH_b$ after several scanning cycles even at 80 K (See **Figure S1** in Supporting Online Materials). We attribute such dissociation of water at the $BBO_V$s to the tip-induced effect at 80 K, as that of the adsorbed molecular oxygen at the $BBO_V$s *(18,19)*.

Whereas, the adsorbed water molecules at the $Ti_{5c}$ sites can be much stable against the tip-induced dissociation up to a threshold bias voltage of about 2.4 V, as shown in **Fig. 1G-1I**. It is found that for the water molecules at the $Ti_{5c}$ sites, one of the hydrogen atoms can be removed under the voltage pulse of 2.4 V, leaving a noisy spot at the $Ti_{5c}$ site (the left panel of **Fig. 1H** and the process 1 in **Fig. 1J**). Both of the hydrogen atoms can be even cut off under a higher voltage pulse of 2.8 V, producing an O adatom at the $Ti_{5c}$ site *(18,19)* (the right panel of **Fig. 1H** and the process 2 in **Fig. 1J**). The noisy spot is assigned to hydroxyl at the $Ti_{5c}$ site, $OH_t$, which can be further dissociated to an O adatom (**Fig. 1I** and the process 3 in **Fig. 1J**). These operations are a result of the inelastic electron tunneling, similar to what was observed before for the tip-induced desorption of hydrogen from $OH_b$ *(20)*. To avoid such tip-induced dissociation, the measurements presented in the following part were performed at around 1.0 V and 10 pA if not specified.

**Fig. 2** shows the direct evidence that the adsorbed molecules at $Ti_{5c}$ sites can be dissociated under UV

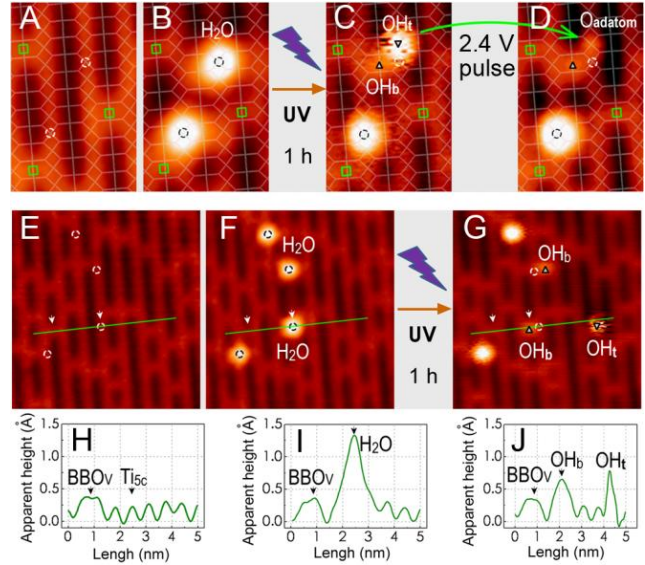

**FIG. 2.** Photocatalytic dissociation of single water molecules. (A) and (B) STM images (Size: 1.9×2.9 $nm^2$) before and after water adsorption. (C) Image after 266 nm UV irradiation for 1 h. (D) Image showing the further dissociation of the noisy $OH_t$ to an oxygen adatom at the $Ti_{5c}$ site by applying a voltage pulse of 2.4 V. (E)-(G) Another set of images (6.3×6.6 $nm^2$) showing the photocatalytic dissociation of water molecules under 400 nm UV irradiation for 1 h. (H)-(J) Line profiles along the lines in (E), (F), and (G), respectively. Circles: adsorption sites of water at $Ti_{5c}$, rectangles: $BBO_V$ sites, upward triangles: $OH_b$, downward triangles: $OH_t$. Imaging conditions: 1.0 V and 10 pA, 80 K.

irradiation. In the set of **Fig. 2A-2C**, it can be seen that one of the water molecules disappears from its original $Ti_{5c}$ site after 266 nm UV irradiation, accompanying with the occurrence of a less protruded spot at the adjacent oxygen of the BBO row and a noisy spot at a nearby $Ti_{5c}$ site (from the original water adsorption site by one lattice distance). The noisy spot resembles well the $OH_t$ in **Fig. 1H**, and the spot at the adjacent bridge oxygen behaves as $OH_b$. This nicely demonstrates that the water can indeed be dissociated into an $OH_t$ and a hydrogen atom, while the latter transfers to the adjacent BBO to form $OH_b$. To confirm that the noisy spot corresponds to $OH_t$, we applied a voltage pulse of 2.4 V on it and obtained an O adatom, as expected in **Fig. 2D**.

Similar results were obtained under irradiation of UV light with different wavelengths. **Fig. 2E-2G** show another set of images under irradiation of 400 nm UV light. It is observed that $OH_b$ always presents at the adjacent bridge oxygen with water dissociation under UV irradiation, as shown in **Fig. 2G**, which differs from the tip-induced water dissociation (**Fig. 1H** and **1I**). Two water molecules dissociate in the frame of **Fig. 2G**, but only one OH$_t$ can be observed at a Ti$_{5c}$ site away from the original place by several lattice distance over the BBO rows, implying that OH$_t$ is quite diffusive, at least under UV irradiation. This finding is consistent with the LIF results *(14)*. The line profiles shown in **Fig. 2H-2J** give the apparent height of these different species, which have also been further confirmed by the tip manipulation (See **Figure S2** in Supporting Online Materials).

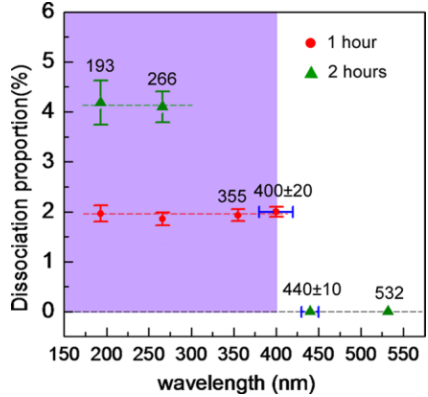

**FIG. 3.** Photodissociation proportion of water at Ti$_{5c}$ versus light wavelength for different irradiation times. The vertical error bars give the standard deviation of the data from more than three runs. The horizontal bars for 400 and 440 nm indicate the bandwith of the filters when the mercury-xenon lamp was used. Different light sources were used: mercury-xenon lamp for 400 (with bandwidth 40 nm, 0.1 mW/cm$^2$) and 440 (with bandwidth 20 nm, 0.1 mW/cm$^2$), Nd:YAG laser for 532, 355, and 266 nm (1~10 mW/cm$^2$), and excimer laser for 193 nm (1~10 mW/cm$^2$).

We also examined the effect of the wavelength and the intensity of lights on the dissociation probability. As shown in **Fig. 3**, the dissociation events can be observed only when the wavelength of light is shorter than 400 nm, which energy accords well to the band gap 3.1 eV of the rutile TiO$_2$. It is noticed that although we do observe the photocatalytic water dissociation, after counting thousands of adsorbed water molecules, we only obtained about 2% and 4% dissociation events for 1 and 2 h UV irradiation, respectively. It is also difficult to observe the dissociation event for short irradiation time less than 30 min. Furthermore, the observed dissociation probability is not obviously dependent on the wavelength (400 ~ 193 nm), nor on the light intensity (0.1~10 mW/cm$^2$). The independence of dissociation probability on the light intensity is consistent with the suggestion that the photocatalytic processes are limited to a low-intensity of UV irradiation *(21,22)*. As a comparison, we performed the experiment for co-adsorbed water and methanol under UV irradiation (See **Figure S3** in Supporting Online Materials), revealing that the dissociation of water is more difficult than that of methanol *(23)* and ethanol *(24)*.

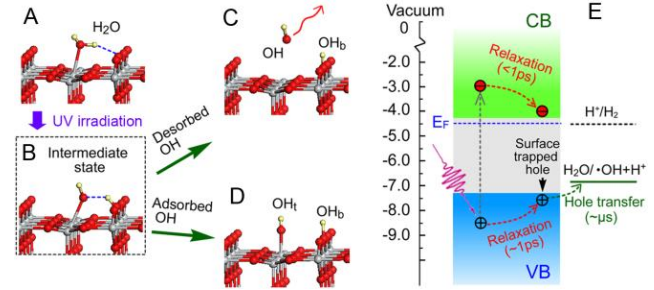

**FIG. 4.** Illustration of photocatalytic dissociation of water. (A) Structural model of water at a Ti$_{5c}$ site. (B) Proton transfer due to the reaction of the water molecule with the trapped hole. (C) Dissociation of the adsorbed water molecule into OH$_b$ and desorbed •OH, or (D) OH$_b$ and adsorbed OH$_t$ at Ti$_{5c}$. (E) Energy diagram and relaxation time scales of the photogenerated electron-hole pair (Refs. *8*, *25-29*) .

On the basis of our observations, we suggest that the initial reaction step should simply follow photooxidation process as

$$H_2O + h^+ \rightarrow \bullet OH + H^+ \qquad (1)$$

where h$^+$ refers the VB hole. This fits well with the suggestion of Henderson that water photoreduction on rutile TiO$_2$(110) does not proceed via electron attachment to adsorbed water molecules *(2)*. The mechanism for the observed photodissociation of single water molecules under UV irradiation is schematically illustrated in **Fig. 4**. It shows that in this reaction the proton transferring from the water molecule to the adjacent bridge oxygen always happens (**Fig. 4B**) once the trapped hole reacts with the adsorbed water molecule, and the produced •OH species may either desorb from the surface or adsorb at a certain Ti$_{5c}$ site (**Fig. 4C** and **4D**). It is known that the photo-generated hole-electron pairs can be quickly relaxed to their equilibria (**Fig. 4E**). The relaxation of deep holes to the VB maximum at the surface is in the time scale of a few ps *(2,25,26)*, while the transfer of the surface trapped holes to the adsorbed water molecules is around several μs *(27)*. In comparison, the relaxation of hot electrons is in the scale of several tens to hundred fs *(28,29)*, while the transfer of the trapped electrons to the adsorbed water molecules is in the much longer time scale of several tens to hundreds µs *(27)*. Hence, the reaction of the holes with the water molecules should be the dominant factor. This may also explain why the dissociation process is much difficult to take place through the electron attachment. The fast relaxation of the holes and the electrons results in the observed independence of UV wavelength on dissociation probability.

To conclude, we have provided the direct experimental evidence that water molecules on $TiO_2$ surface can be photocatalytically dissociated to form hydroxyls. The discovery of this important initial reaction step in the photocatalytic water splitting reaction chain is significant. It not only contributes greatly to the fundamental understanding of the water splitting reaction on $TiO_2$, but also provides a new strategy for the design of better systems for efficient water splitting reaction in general.

**Acknowledgements**: This work was supported by NBRP (grants 2011CB921400 and 2010CB923300) and NSFC (grants 9021013, 10825415, 10874164, 21003113, 21121003), China.

# Supporting Materials

**Evidence of Photocatalytic Dissociation of Water on TiO$_2$ with Atomic Resolution**

Shijing Tan, Yongfei Ji, Yang Wang, Jin Zhao, Aidi Zhao, Bing Wang*, Yi Luo, Jinlong Yang, and Jianguo Hou*

Hefei National Laboratory for Physical Sciences at the Microscale, University of Science and Technology of China, Hefei, Anhui 230026, P.R. China (bwang@ustc.edu.cn and jghou@ustc.edu.cn)

## Materials and Methods

Our STM experiments were conducted with a low temperature scanning tunneling microscope (Matrix, Omicron) in an ultra-high vacuum system with a base pressure less than $3\times10^{-11}$ Torr. The STM measurements were mainly performed at 80 K. An electrochemically etched polycrystalline tungsten tip was used in STM experiments. The rutile TiO$_2$ (110) sample (Princeton Scientific Corporation) was prepared by repeated cycles of ion sputtering (3000 eV Ar$^+$) and annealing to 900 K with a Ta-foil heater behind the sample. Water (Aldrich, deuterium-depleted, 99.99%) was purified by several freeze-pump-thaw cycles using liquid nitrogen. Water was transferred directly to the TiO$_2$ surface through a dedicated tube in the chamber. The outlet of the tube was only about 5 mm from the TiO$_2$ surface. This method can minimize the background water which may possibly cause misleading. During water dosing, the sample was maintained at 80 K. The light irradiation experiments were performed with different wavelengths by using light sources of Mercury-xenon lamp (Hammatsu, L2423, with bandpass filters: centered at 400 and 440 nm with bandwidths of 40 and 20 nm, respectively, typical intensity: 0.1 mW/cm$^2$), Nd:YAG laser (Spectra-Phyiscs, Pro-250, repetition: 10 Hz, duration: 10 ns, for wavelength of 532, 355, and 266 nm, intensity: 1~10 mW/cm$^2$), and Excimer laser (Coherent Inc., COMPexPro 201, ArF, repetition: 4 Hz, duration: 20 ns, for wavelength of 193 nm, intensity: 1~10 mW/cm$^2$). A specially coated sapphire window was used for 193 nm UV light, which allows 90% transmission of 193 nm UV light. During light illumination, the tip was retracted back for about 10 μm to prevent from shadowing effect. We always compared the areas before and after light irradiation to trace any change of the surface *(23)*.

## 1. The different behaviors between the dissociation of water at the BBO$_V$ sites and the photocatalytic dissociation of water at the Ti$_{5c}$ sites

We have compared the behavior of the spontaneous dissociation of water at the BBO$_V$ sites with that of photocatalytic dissociation of water at the Ti$_{5c}$ sites at 80 K, as shown in **Figure S1**. It is observed that the water molecules at the BBO$_V$ sites may dissociate before UV irradiation. As shown in the marked rectangle I in **Figure S1-B** and **S1-C** (and see the corresponding magnified images and line profiles in upper-right panel), one of the water molecules at the BBO$_V$ undergo dissociation after consecutive scanning, forming a pair of OH$_b$. At elevated temperatures, water dissociation at the BBO$_V$ sites has been widely observed before *(12,13,30)*, which is a thermal-driven spontaneous process. At 80 K, such dissociation is more likely to be tip-induced, as that of molecular oxygen at the BBO$_V$ sites *(18,19,31)*.

Under UV irradiation, as also shown in the main text, we observed the dissociation of water at the Ti$_{5c}$ sites (**Figure S1-D** and the magnified images and line profiles in the lower-right panel). The photocatalytic dissociation of water always produces a hydroxyl at the adjacent bridge oxygen, due to the product of the hydrogen atom bonding to the bridge oxygen, while the •OH (may adsorb as OH$_t$, see the description in the main text) diffuses away or even desorbs from the surface. It is noticed that the water molecules at the $BBO_V$ sites do not show any dissociation priority under UV irradiation. Furthermore, the formed $OH_b$ pair is not separated under UV irradiation, unlike the thermal-driven diffusion and separation of $OH_b$ pair at elevated temperatures *(13)*. This fact also suggests that $OH_b$ is not as sensitive as $OH_t$ under UV irradiation. In a real reaction environment, once a $BBO_V$ site is filled with a dissociative water molecule, $OH_b$ pair is consequently formed. Less $BBO_V$ sites are available, which should result in the reduction of this activity. Such a drawback could be overcome by the direct photocatalytic dissociation of water.

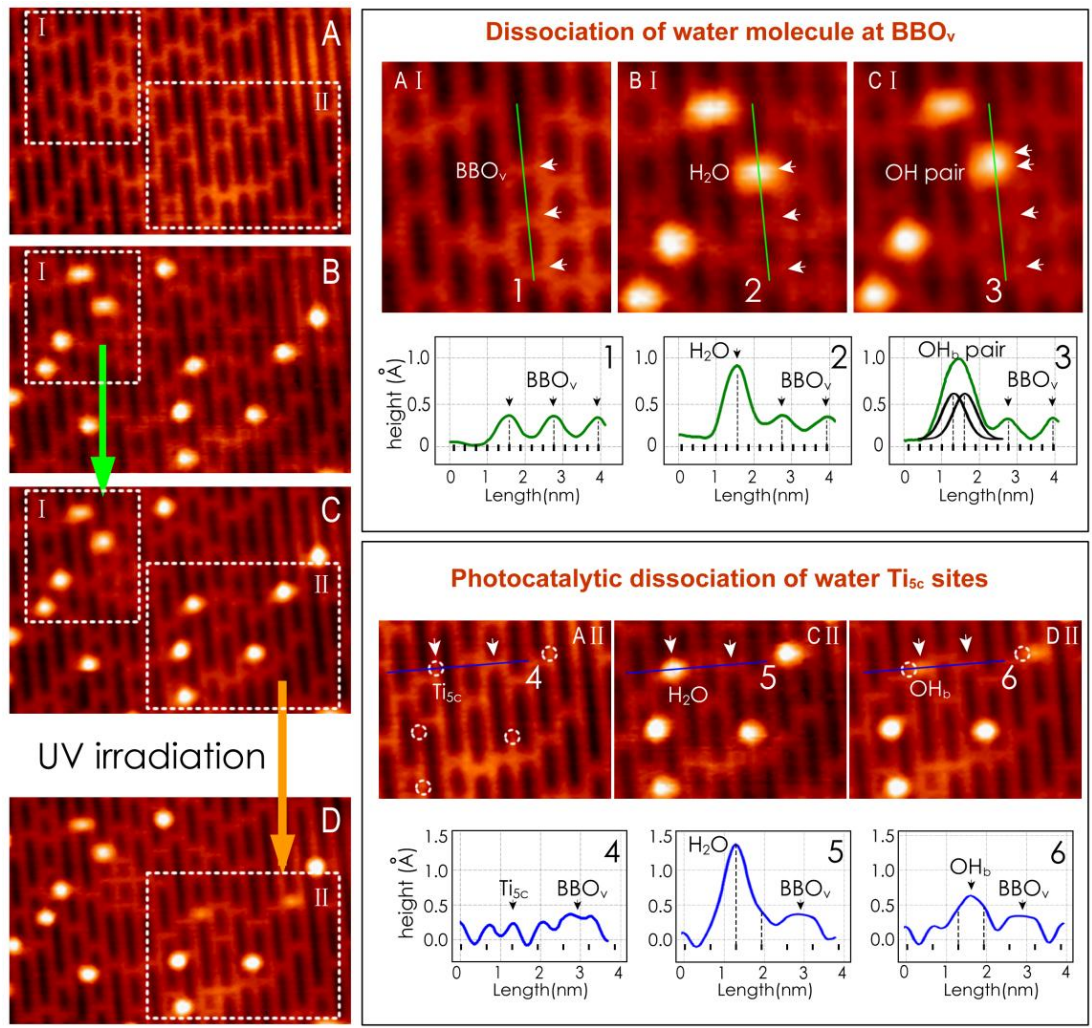

**Figure S1:** Comparing the dissociation of water at $BBO_V$ sites and the photocatalytic dissociation of water at $Ti_{5c}$ sites. (A) STM image of clean surface. (B) Image after water dosing. (C) Consecutively acquired image. (D) Image ffter 266nm UV irradiation for 1 hour. Size: 11.7×7.7 $nm^2$. Imaging conditions: 1.0 V and 10 pA. The magnified images of (A-I) to (C-I) and the corresponding line profiles showing dissociation of one water molecule at $BBO_V$ to an $OH_b$ pair. Size: 4.4×4.8 $nm^2$. The magnified images of (A-II), (C-II), (D-II) and the corresponding line profiles showing that dissociation of two water molecules at $Ti_{5c}$ sites after UV irradiation. Size: 6.5×5.0 $nm^2$.

## 2. Confirmation of the reaction products from photocatalytic dissociation of water.

In **Fig. 2G** in the main text, we attributed the new spots at BBO as hydroxyls ($OH_b$) and the noisy spot at $Ti_{5c}$ site as $OH_t$, the products from the photocatalytic dissociation of water. Here we show further experimental evidence to confirm our assignments. **Figure S2-A** is the same as **Fig.2G** in the main text, which shows the changes of the dissociation of two water molecules under UV irradiation. First, a 2.4 V pulse was applied on the $OH_t$, resulted in the noisy $OH_t$ changing to O adatom ($O_a$) by removing the hydrogen atom, as shown in **Figure S2-B**. Followed, we have made another two pulses of 2.8 V on the $OH_b$s, respectively, removing the hydrogen atoms from both of the $OH_b$s, as shown in **Figure S2-C**. The whole process could confirm our assignments of $OH_b$ and $OH_t$ affirmatively.

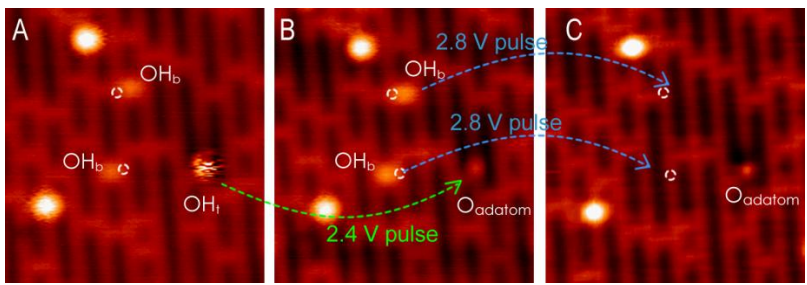

**Figure S2:** (A) STM image with photocatalyically dissociated products of $OH_b$ and $OH_t$ (the same image as shown in **Fig. 2G** in the main text). (B) Image showing the dissociation of $OH_t$ to an oxygen adatom at the $Ti_{5c}$ site by applying a voltage pulse of 2.4 V. (C) Image showing the dissociation of $OH_b$s by applying a voltage pulse of 2.8 V, respectively. Size: $6.3 \times 6.6$ nm$^2$.

## 3. Comparison the photocatalytic dissociation efficiency between water and methanol.

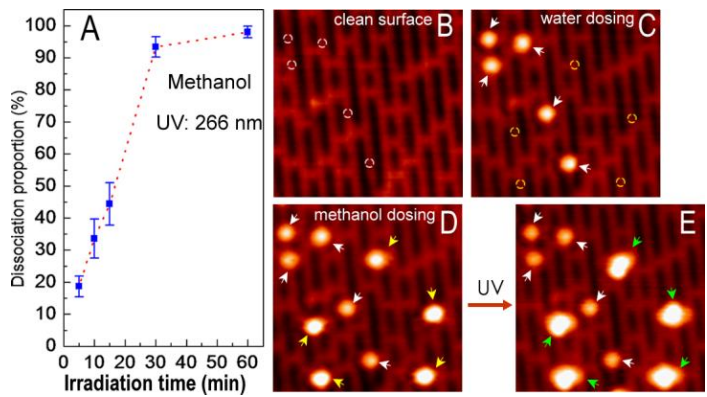

**Figure S3:** (A) Dissociation proportion of methanol under 266 nm UV irradiation as a function of irradiation time. (B) Image of clean $TiO_2$(110)-1×1 surface. (C) Image after water dosing, and (D) methanol dosing followed within the same area, obtaining a co-adsorbed water and methanol sample. (E) Image after 266 nm UV irradiation for 30 min. White arrows: molecular water; yellow arrows: molecular methanol; green arrows: dissociated methanol. Image size: $7.3 \times 7.2$ nm$^2$, 1.0 V and 10 pA, 80 K.

In our previous work, we reported the photocatalytic dissociation of methanol at $Ti_{5c}$ via O-H bond cleavage and the H transferring to adjacent $O_b$ *(23)*. **Figure S3-A** shows the dissociation proportions of methanol as the function of UV irradiation time. The data are obtained under 266 nm UV irradiation. After irradiation for only 5 min, ~18% methanol molecules are dissociated. The dissociation proportion increases rapidly with the increase of the UV irradiation time from 15 to 30 min, going from ~33% to ~90%. We can observe that under UV irradiation for 60 min, nearly all the methanol molecules could be dissociated. However, water has low dissociation probabilities, as we observed (**Fig. 3** in the main text), only ~2% water dissociates under UV irradiation for 60 min and increased to ~4% for 120 min. These results show that about two orders of magnitude difference of the photocatalytic reaction probability between methanol and water.

**Figure S3-B** to **S3-E** is a set of STM experiment directly illustrates the difference in the dissociation proportion between methanol and water. We conducted the co-adsorption of water and methanol in the same sample. **Figure**

**S3-B** is the original clean surface. **Figure S3-C** shows the image after water dosing, and **Figure S3-D** shows the image after methanol dosing followed with similar concentration of about 0.02 ML. The methanol molecules are more protruded than the water molecules (**Figure S3-D**). After 266 nm UV irradiation for 30 min, we can see that all of the five methanol molecules are dissociated *(23)*, but all of the water molecules remain unchanged, as shown in **Figure S3-E**.